\documentclass[smus]{snow2e}
\usepackage{epsf}
\begin{document}

\title{Hybrid Rings of Fixed 8T Superconducting Magnets and
       Iron Magnets Rapidly Cycling between -2T and +2T for a Muon
       Collider
       \thanks{Work supported by U.~S.~DOE \ DE-FG05-91ER40622
       and DE-AC02-76CH00016}}

\author{D. J. Summers\\ {\it Department of Physics and Astronomy,
         University of Mississippi--Oxford, University, MS 38677 \ USA}}

\maketitle

\vspace*{-3.8cm}
\leftline{
Proceedings, \, New Directions for High Energy Physics, \, 
Amer.~Phys.~Soc., \, Snowmass, CO \, (25 June -- 12 July 1996) \, 238}
\vspace*{2.6cm}

\setcounter{page}{238}

\begin{abstract} 
Two 2200\,m radius hybrid rings of fixed superconducting magnets
and iron magnets 
ramping at 200 Hz
and 330 Hz are used to accelerate muons.  Muons are given 25 GeV of 
RF energy per orbit.  Acceleration is from 250 GeV/c to 2400 GeV/c 
and requires a total of 86 orbits in both rings; 82\% of the muons survive.
The total power consumption of the iron dipoles is 4 megawatts. Stranded
copper conductors 
and thin Metglas laminations are used to reduce power losses.   
\end{abstract}

\section{INTRODUCTION}

   For a $\mu^+\mu^-$ collider, muons must be rapidly accelerated to high
energies while minimizing the kilometers of radio frequency (RF) cavities and 
magnet bores.  Cost must be moderate. Some muons may be lost to
decay but not too many.

   Consider a ring of fixed superconducting magnets alternating with
iron magnets rapidly cycling between full negative and full positive
field [1].  Table I shows the range of average dipole magnetic field for
various mixes of the two types of magnets.  One might use more than one
ring in succession.  Now proceed with a few {\it back--of--the--envelope}
calculations.

\begin{table}[bh]
\begin{center}
\caption{Hybrid ring parameters.}
\renewcommand{\arraystretch}{1.1}
\begin{tabular}{cccc} \hline \hline
8T      &   $\pm$2T   &   Initial &  Final    \\ 
Magnets &   Magnets &   B Field &  B Field  \\ \hline
22\%    &   78\%    &   0.2T    &  3.3T     \\           
25\%    &   75\%    &   0.5T    &  3.5T     \\ 
35\%    &   65\%    &   1.5T    &  4.1T     \\ 
40\%    &   60\%    &   2.0T    &  4.4T     \\ 
50\%    &   50\%    &   3.0T    &  5.0T     \\ 
52\%    &   48\%    &   3.2T    &  5.1T     \\ 
55\%    &   45\%    &   3.5T    &  5.3T     \\ 
60\%    &   40\%    &   4.0T    &  5.6T     \\ 
70\%    &   30\%    &   5.0T    &  6.0T     \\ 
80\%    &   20\%    &   6.0T    &  6.8T     \\ \hline \hline
\end{tabular}
\end{center}
\end{table}

\section{MAGNET SAGITTAS}

   The sagitta of a muon in a magnet increases linearly with increasing
magnetic field, $B$. It decreases linearly with increasing momentum, $p$.  And
it increases as the square of the length of a magnet, $\ell$.  The size of the
sagitta directly affects the size of magnet bores because the sagitta {\it
changes} throughout a cycle. Table II shows sagitta for various magnets and
momenta.  As momentum increases, the sagitta in the 8 Tesla magnets 
decreases towards
zero and the sagitta in the 2 Tesla magnets goes somewhat past zero.  Note
that for a given bore size the magnets can be longer given a higher injection
momentum.

\begin{equation}
\hbox{Sagitta} = R - \sqrt{R^2 - ({\ell}/2)^2}; \quad R = {p \over {.3 B}} 
\end{equation}

\begin{table}[h]
\begin{center}
\caption{Sagitta 
as a function of momentum, magnetic field, and magnet length.}
\renewcommand{\arraystretch}{1.1}
\begin{tabular}{cccc} \hline \hline
Momentum &    B Field &  Length   &  Sagitta \cr
 (GeV)   &    (Tesla) &  (meters) &  (mm)    \\ \hline
250      &    8       &  1.5      &    3           \\ 
250      &    2       &  4.5      &    6           \\ 
250      &    8       &  2        &    5           \\ 
250      &    2       &  6        &   11           \\ 
250      &    8       &  3        &   11           \\ 
250      &    2       &  9        &   24           \\ \hline \hline
\end{tabular}
\end{center}
\end{table}

\section{POWER CONSUMPTION}

   Consider the feasibility of an iron dominated design for a magnet which
cycles from a full -2 Tesla to a full +2 Tesla [2].  First calculate the energy,
$W$, stored in a 2 Tesla field in a volume 6\,m long,\, .03\,m high, and 
.08\,m wide.
The permeability constant, $\mu_0$, is $4\pi\times 10^{-7}$.

\begin{equation}
W = {B^2\over{2{\mu_0}}}[\hbox{Volume}] = 23\,000 \ \hbox{Joules}
\end{equation}

Next given 6 turns, an LC circuit capacitor, and a 250 Hz frequency; estimate 
current, voltage, inductance, and capacitance. The height, $h$, of the 
aperature is\, .03\,m.  
The top and bottom coils may be connected as two separate
circuits to halve the switching voltage.   

\begin{equation}
B = {{\mu_0\,NI}\over{h}}  \quad\rightarrow\quad 
I = {{Bh}\over{\mu_0\,N}} = 8000 \ \hbox{Amps}
\end{equation}
\begin{equation}
W = .5\,L\,I^2  \quad\rightarrow\quad L = {2\,W\over{I^2}} = 
720\,\mu\hbox{H}
\end{equation}
\begin{equation}
f = {1\over{2\pi}}\sqrt{1\over{LC}}  \quad\rightarrow\quad
 C = {1\over{L\,(2\pi f)^2}} = 560\, \mu\hbox{F}
\end{equation}
\begin{equation}
W = .5\,C\,V^2  \quad\rightarrow\quad V = \sqrt{2W\over{C}} = 9000 \ 
\hbox{Volts}
\end{equation}

Now calculate the resistive energy loss, which over time is equal to one-half
the loss at the maximum current of 8000 Amps.  The one-half comes from the 
integral of cosine squared.  
Table III gives the resistivities of copper and other metals.
A six-turn copper conductor 3\,cm thick, 10\,cm high, 
and 7800\,cm long has a power dissipation of 15 kilowatts.

\begin{equation}
R = {7800 \ (1.8\,\mu\Omega\hbox{-cm})\over{(3) \, (10)}} = 470\,\mu\Omega
\end{equation}
\begin{equation}
P = I^2R\int_0^{2\pi}\!\cos^2(\theta)\,d\theta = \hbox{15\,000 watts/magnet} 
\end{equation}

\begin{table}[!htb]
\begin{center}
\caption{Conductor, cooling tube, and 
soft magnetic material properties of resistivity, magnetic saturation in Tesla,
and coercive force in Oersteds [3].}
\renewcommand{\arraystretch}{1.1}
\tabcolsep=0.6mm
\begin{tabular}{llccc} \hline \hline
                   &          &                      & B   &           \\
Material                   & Composition & $\rho$    & Max & H$_c$      \\
                   &             & ($\mu\Omega$-cm) & (Tesla)    & (Oe)  \\
                                                                         \hline
Copper             & Cu                      & 1.8             & --- & ---   \\
Stainless 316L & Fe 70,\, Cr 18,\, Ni 10,    & 74 & --- & ---   \\
               & Mo 2,\, C .03 &  &  &    \\
Stainless 330      & Fe 43,\, Ni 35,\, Cr 19 & 103             & --- & ---   \\
Hastelloy B        & Ni 66,\, Mo 28,\, Fe 5  & 135             & --- & ---   \\
{\em Thermostat} [4]  & Mn 72,\, Cu 18,\, Ni 10  
                                                         & 175 & --- & ---   \\
Thermenol          & Fe 80,\, Al 16, \, Mo 4  & 162            & 0.61 & .02  \\
Pure Iron [5]     & Fe 99.95,\, C .005      & 10              & 2.16 & .05  \\
1008 Steel         & Fe 99,\, C .08           & 12             & 2.09 &  0.8 \\
Grain--Oriented &  Si 3,\, Fe 97           & 47                & 1.95 & .1  \\
Supermendur  [6]  & V 2,\, Fe 49,\, Co 49   & 26              & 2.4  & .2   \\
Hiperco 27   [7]  & Co 27, Fe 71,\, C .01  & 19              & 2.36  & 1.7  \\
Metglas           & Fe 81,\, B 14,\, Si 3, &  135   &  1.6    & .03  \\
\ \ 2605SA1 [8, 9]  & C 2  &     &      &   \\
\hline \hline
\end{tabular}
\end{center}
\end{table}

Calculate the dissipation due to eddy currents in this conductor, which will
consist of transposed strands to reduce this loss [10--12].  
To get an idea, take the maximum B-field
during a cycle to be that generated by a 0.05m radius conductor carrying
24000 amps.  
This ignores fringe fields from the gap which will make the real answer higher.
The eddy current loss in a rectangular conductor made of square wires 
1/2 mm wide with a perpendicular magnetic field is as follows.  
The width of the wire is $w$.

\begin{eqnarray}
B & = & {{\mu_0\,I}\over{2\pi r}} = 0.096 \ \hbox{Tesla}  \\
P & = & \hbox{[Volume]}{{(2\pi\,f\,B\,w)^2}\over{24\rho}}  \\
  & = & [.03 \ .10 \ 78]\, {{(2\pi \ 250 \ .096 \ .0005)^2} \over 
{(24)\,1.8\times{10^{-8}}}} = 3000 \ 
\hbox {watts}
\nonumber
\end{eqnarray}

Cooling water will be needed, so calculate the eddy current losses
for cooling tubes made from type 316L stainless
steel. More exotic metals with higher resistivities are also available as shown 
in Table III. 
Choose 2 tubes per 3\,cm $\times$ 10\,cm 
stranded copper conductor for a total length of 78 $\times$ 2 = 156\,m.  
Take a 12\,mm OD
and a 10\,mm ID. Subtract the losses in the 
inner {\it missing} round conductor.
The combined eddy current loss in the copper plus the stainless steel is
4200 watts (3000 + 2400 - 1200).

\begin{eqnarray}
P(12\,\hbox{mm}) 
 & = & \hbox{[Volume]}\, {{(2\pi\,f\,B\,d)^2} \over {32\,\rho}} \\
 & = & [\pi \ .006^2 \ 156]\,  {(2\pi\, 250 \ .096 \  .012)^2 \over 
{(32) \ 74 \times 10^{-8}}} \nonumber \\ 
 & = & 2400 \ \hbox{watts} \nonumber \\
P(10\,\hbox{mm}) 
& = & \hbox{[Volume]}\, {{(2\pi\,f\,B\,d)^2} \over {32\,\rho}} \\
 &         = & [\pi \ .005^2 \ 156]\, {{(2\pi\,250 \ .096 \ .010)^2}
\over {(32) \ 74{\times}10^{-8}}} \nonumber \\  
& = &  1200 \ \hbox{watts} \nonumber
\end{eqnarray}

   Eddy currents must be reduced in the iron not only because of the increase
in power consumption and cooling, but also because they introduce multipole
moments which destabilize beams.  If the laminations are longitudinal,
it is hard to force the magnetic field to be parallel to the laminations
near the gap.  This leads to additional eddy current gap losses [13].  
So consider a magnet with transverse laminations as sketched in Fig.~1
and calculate the eddy current losses. 
The yoke is either
0.28\,mm thick 3\% grain oriented silicon steel [14--17] 
or 0.025\,mm thick Metglas 2605SA1 [8, 9].
The pole tips are 0.1\,mm thick Supermendur [6] to increase 
the field in the gap [18].

\begin{eqnarray}
\lefteqn{\hbox{P(3\% Si--Fe)} 
=  \hbox{[Volume]}{{(2\pi\,f\,B\,t)^2}\over{24\rho}}} \\
& &   =  [6 \, ((.42 \ .35) - (.20 \ .23))]\, 
{{(2\pi \ 250 \ 1.6 \ .00028)^2} \over 
{(24)\,47\times{10^{-8}}}}  \nonumber \\ 
& & =  27\,000 \ 
\hbox {watts}
\nonumber
\end{eqnarray}

\begin{eqnarray}
\lefteqn{\hbox{P(Metglas)}  
=  \hbox{[Volume]}{{(2\pi\,f\,B\,t)^2}\over{24\rho}}} \\
& &   =  [6 \, ((.42 \ .35) - (.20 \ .23))]\, 
{{(2\pi \ 250 \ 1.6 \ .000025)^2} \over 
{(24)\,135\times{10^{-8}}}}  \nonumber \\ 
& & =  75 \ 
\hbox {watts}  
\nonumber
\end{eqnarray}

\begin{eqnarray}
\lefteqn{\hbox{P(Supermendur)} 
= \hbox{[Volume]}{{(2\pi\,f\,B\,t)^2}\over{24\rho}}} \\
& &   =  [6 \, \, .09 \, \, .02]\, 
{{(2\pi \ 250 \ 2.2 \ .0001)^2} \over 
{(24)\,26\times{10^{-8}}}}  \nonumber \\ 
& & =  210 \ 
\hbox {watts} 
\nonumber
\end{eqnarray}

\begin{figure}[htb]
\begin{center}
\vspace*{-2.0cm}
\epsfysize=11.8cm\epsffile{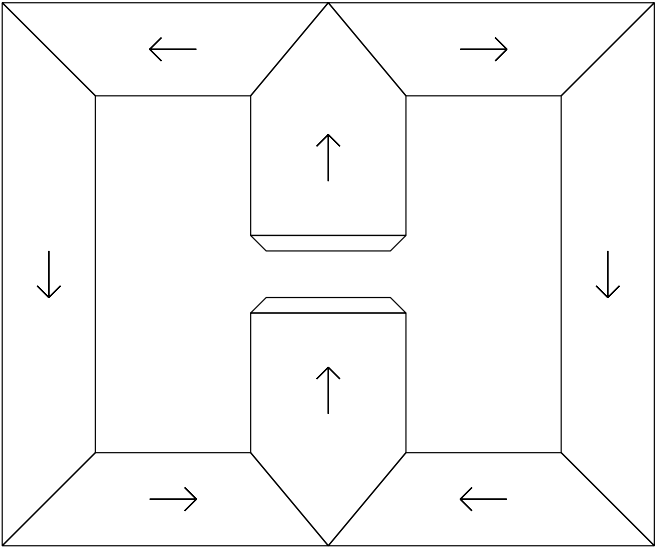} 
\parbox{8.9cm}{\caption{
A two dimensional picture of an H frame magnet lamination with grain 
oriented 3\%\,Si--Fe steel.  The 
arrows show both the magnetic field direction and the grain direction of 
the steel. Multiple pieces are used to exploit the high permeability and 
low hysteresis in the grain direction [19].
If Metglas 2605SA1 is used for the yoke, multiple pieces are not needed, except
for the poles.
The pole tips are an iron--cobalt 
alloy for flux concentration exceeding 2 Tesla.}}
\end{center}
\end{figure}

   Eddy currents are not the only losses in the iron.  Hysteresis 
losses,
$\int{\bf{H}}{\cdot}d\,{\bf{B}}$, scale  
with the coercive force,
H$_c$, and increase linearly with frequency.
Anomalous loss [5] which is difficult to calculate
theoretically must be included.  Thus I now use functions fitted 
to experimental measurements of 0.28\,mm thick 3\% grain oriented 
silicon steel [20], 
0.025\,mm thick Metglas 2605SA1 [8],
and 0.1\,mm thick Supermendur [20].

\begin{table}[!htb]
\begin{center}
\caption{Magnet core materials.}
\renewcommand{\arraystretch}{1.1}
\tabcolsep=1.0mm
\begin{tabular}{lcccc} \hline \hline
Material            & Thickness   & Density     & Volume     & Mass \\
                    & (mm)        & (kg/m$^3$)  & (m$^3$)    & (kg) \\ \hline
3\% Si--Fe & 0.28   & 7650        & 0.6        & 4600 \\
Metglas             & 0.025       & 7320        & 0.6        & 4400 \\
Supermendur         & 0.1         & 8150        & 0.01       & 90   \\
\hline \hline
\end{tabular}
\end{center}
\end{table}

\begin{eqnarray}
\hbox{P(3\% Si--Fe)} & = & 4.38 \times 10^{-4} \, f^{1.67} \, B^{1.87}    \\   
\nonumber            & = & 4.38 \times 10^{-4} \, 250^{1.67} \, 1.6^{1.87} \\ 
\nonumber            & = & 10.7 \, \, \hbox{watts/kg}                      \\
\nonumber            & = & 49\,000 \, \, \hbox{watts/magnet} 
\end{eqnarray}

\begin{eqnarray}
\hbox{P(Metglas)}    & = & 1.9 \times 10^{-4} \, f^{1.51} \, B^{1.74}    \\   
\nonumber            & = & 1.9 \times 10^{-4} \, 250^{1.51} \, 1.6^{1.74} \\ 
\nonumber            & = & 1.8 \, \, \hbox{watts/kg}                         \\
\nonumber            & = & 7900 \, \, \hbox{watts/magnet}  
\end{eqnarray}

\begin{eqnarray}
\hbox{P(Supermendur)} & = & 5.64 \times 10^{-3} \, f^{1.27} \, B^{1.36}    \\   
\nonumber            & = & 5.64 \times 10^{-3} \, 250^{1.27} \, 2.2^{1.36} \\ 
\nonumber            & = & 18 \, \, \hbox{watts/kg}                         \\
\nonumber            & = & 1600 \, \, \hbox{watts/magnet}
\end{eqnarray}

\begin{table}[!hbt]
\begin{center}
\caption{Power consumption for a 250 Hz dipole magnet.}
\renewcommand{\arraystretch}{1.1}
\begin{tabular}{lcc} \hline \hline
Material                   &    3\% Si--Fe     &       Metglas    \\       
Coil Resistive Loss    & 15\,000 watts &      15\,000 watts \\
Coil Eddy Current Loss &  4200 watts      &      4200 watts  \\
Core Eddy Current Loss      &  27\,210  watts      &      285 watts   \\ 
Total Core Loss             &  50\,600 watts        &      9500 watts \\ \hline
Total Loss                  &  69\,800 watts &  28\,700 watts  \\ \hline \hline
\end{tabular}
\end{center}
\end{table}

   In summary, a 250 Hz dipole magnet close to 2 Tesla looks possible as long
as the field volume is limited and one is
willing to deal with stranded copper and thin, low hysteresis laminations.
Total losses can be held to twice the I$^2$R loss in the copper alone, 
using Metglas.

\section{MUON ACCELERATION AND SURVIVAL}

   Now with a rough design for a fast ramping magnet in hand, work out the
details of ring radii, RF requirements, and the fraction of muons
that survive decay. The fraction of the circumference packed with dipoles is
set at $P_F$ = 70\%.  As an example, consider two rings in a 2200\,m radius 
tunnel with an injection momentum of 250 GeV/c.  The first has 25\%
8T magnets and 75\% $\pm$2T magnets and ramps from 0.5T to 3.5T.
The second has 55\%
8T magnets and 45\% $\pm$2T magnets and ramps from 3.5T to 5.3T.

\begin{equation}
B = {{250\,\hbox{GeV/c}} \over {.3\,P_F\,R}} =  
{{250} \over {(.3)\,(.7)\,(2200)}} = 0.54\,\hbox{Tesla}
\end{equation}

\begin{eqnarray}
p & = & (3.5\,\hbox{Tesla})\,(.3)\,(P_F)\,(R) \\ 
  & = & (3.5)\,(.3)\,(.7)\,(2200) = 1600\,\hbox{GeV/c} \nonumber 
\end{eqnarray}

\begin{eqnarray}
p & = & (5.3\,\hbox{Tesla})\,(.3)\,(P_F)\,(R) \\ 
  & = & (5.3)\,(.3)\,(.7)\,(2200) = 2400\,\hbox{GeV/c} \nonumber 
\end{eqnarray}

Provide 25 GeV of RF.
The first ring accelerates muons from 250 GeV/c to 1600 GeV/c in 54 orbits.
The second ring accelerates muons from 1600 GeV/c to 2400 GeV/c in 32 orbits.
At what frequency do the two rings have to ramp?

\begin{eqnarray}
\hbox{Time}\,(0.5T \rightarrow 3.5T) & = &  {{(54)\,(2\pi)\,(2.2)} \over 
{300\,000}} \\
& = & 2.5\,\hbox{ms} \nonumber \\ 
& \rightarrow &  200\, \hbox{Hz} \nonumber
\end{eqnarray}

\begin{eqnarray}
\hbox{Time}\,(3.5T \rightarrow 5.3T) &  = & {{(32)\,(2\pi)\,(2.2)} \over 
{300\,000}} \\
& = & 1.5\,\hbox{ms} \nonumber \\
& \rightarrow & 330\, \hbox{Hz} \nonumber
\end{eqnarray}

How many muons survive during the 86 orbits from 250 GeV/c to 2400 GeV/c?
$N$ is the orbit number,
$\tau = 2.2\times10^{-6}$ is the muon lifetime, and $m = .106$ GeV/c$^2$
is the muon mass.

\begin{equation}
\hbox{SURVIVAL} = \prod_{N=1}^{86} \exp\left[{{-2\pi{R}\,m} \over 
{[250 + (25\,N)]\,c\tau}}\right] = 82\%
\end{equation}

Only 1/6 of the 18\% loss occurs in the second ring, so it is not crucial
to run it as fast as 330 Hz; but the RF does allow this speed. 

\section{SUMMARY}

The 250 $\rightarrow$ 1600 GeV/c ring has 1200 6\,m long dipole magnets
ramping at 200 Hz. The 1600 $\rightarrow$ 2400 GeV/c ring has 725 6\,m long
dipole magnets ramping at 330 Hz.  The weighted average rate is 250 Hz.  If
running continuously, the 1925 magnets would consume a weighted average of 29
kilowatts each for a total of 56 megawatts. But given a 15 Hz refresh rate for
the final muon storage ring [21], 
the average duty cycle for the 250 $\rightarrow$
2400 GeV/c acceleration rings is 6\%. So the power falls to 4 megawatts,
which is small.

Finally note that 
one can do a bit better than 82\% on the muon survival during final
acceleration if the first ring is
smaller, say 1000 meters, rather than 2200 meters.  Given that RF is expensive,
a single line of cavities could still be used for all rings.

I would like to thank  K.~Bourkland, 
R.~Fernow, J.~Gallardo, C.~Johnstone, H.~Kirk,
D.~Neuffer, A.~Otter, R.~Palmer, A.~Tollestrup, K.~Tuohy, D.~Walz,
R.~Weggel, E.~Willen, and D.~Winn
for their help and suggestions.


\begin{thebibliography}{99}

\bibitem{sesaps} 
      D.~J.~Summers, {\it The Top Quark, the Higgs Boson, and Supersymmetry at
      $\mu^+ \mu^-$ Colliders,} Invited Talk at the 61st Meeting of the
      Southeastern Section APS, Newport News, Virginia
      (10--13 November 1994); \ Bull.~APS 39 (1994) 1818.

\bibitem{resnick}  
    David Halliday and Robert Resnick, {\it Physics for Students of Science and
      Engineering,} Part II, 2nd Edition, Wiley (1962).

\bibitem{handbook} 
      {\it Metals Handbook, Properties and Selection of Metals,}
      Vol.~1, 8th Edition, 
      American Society for Metals 
      (1961).

\bibitem{fine} 
      California Fine Wire Company, 338 South Fourth Street, 
      Grover Beach, CA 93433.

\bibitem{chen} 
      Chih-Wen Chen, {\it Magnetism and Metalurgy of Soft Magnetic Materials,}
      Dover Publications (1986).

\bibitem{arnold}
       Arnold Engineering Company, 300 North West Street, Marengo, IL 60152.

\bibitem{carp}
        Carpenter Technology, 101 West Bern Street,  Reading, PA 19603.

\bibitem{allied} 
        Allied Signal, Amorphous Metals Division, 
        6 Eastmans Road, Parsippany, NJ 07054. 

\bibitem{spang}
        Magnetics, Division of Spang \& Company, 
        900 East Butler Road, Butler, PA 16003.

\bibitem{sasaki}
      H.~Sasaki, {\it Magnets for Fast--Cycling Synchrotrons,} 
      Invited talk at International Conference on Synchrotron
      Radiation Sources, Centre for Advanced Technology, Indore,
      India (3--6 February 1992); \
      KEK 91-216 (March 1992).

\bibitem{otter89}
      A.~J.~Otter, {\it A Prototype for the Booster Dipole,}
      TRIUMF, TRI-DN-89-K18 
      (February 1989).

\bibitem{otter88} 
      Alan J.~Otter, {\it LAMPF Type Stranded Magnet Cable,}
      TRIUMF, TRI-DN-88-K12 
      (December 1988).

\bibitem{walz}
      W.~O.~Brunk and D.~R.~Walz, {\it A New Pulse Magnet Design Utilizing Tape
      Wound Cores,} 7th Natl.\ Part.\ Accel.\ Conf., 
      Washington (12--14 March 1975) 1548;
      SLAC--PUB--1551 (March 1975) 4 pages.

\bibitem{ludlum}
      Allegheny Ludlum, 1000 Six PPG Place, Pittsburgh, PA 15222. 

\bibitem{armco}
       Armco, Box 1609, Butler, PA 16001.

\bibitem{kawasaki}
        Kawasaki Steel, 55 E.~52nd St., New York, NY 10055.

\bibitem{nippon}
        Nippon Steel, 10 E.~50th St., New York, NY 10022.

\bibitem{pole}
      R.~A.~Early, J.~K.~Cobb, and J.E.~Oijala, {\it Design Calculations and 
      Measurements of a Dipole Magnet with Permendur Pole Pieces,} 
      IEEE Part.~Accel. (1989) 351; \ SLAC-PUB-4883 (March 1989) 3~pages.

\bibitem{grain}
      P.~Schwandt, {\it Comparison of Realistic Core Losses in the Booster Ring
      Dipole Magnets for Grain--Oriented and Ordinary Lamination Steels,}
      TRIUMF, TRI--DN--89--K31 (April 1989).

\bibitem{core}
       Wm.~T.~McLyman, {\it Magnetic Core Selection for 
      Transformers and Inductors,}
      Marcel Decker, ISBN 0-8247-1873-9 (1982). 

\bibitem{juan}
      {\it $\mu^+\mu^-$ Collider: A Feasibility Study}, Juan C.~Gallardo,
      editor, BNL--52503, Fermi Lab--Conf.--96/092, LBNL-38946 (18 June 1996).

\end{thebibliography}
\end{document}